\documentclass[12pt]{article}
\usepackage{amsmath}
\usepackage{amsfonts}
\usepackage{graphicx,epsfig}

\newcommand{\cfield}{\mathbb{C}}
\newcommand{\ee}{{\mathbf e}}
\newcommand{\ff}{{\mathbf f}}
\newcommand{\hh}{{\mathcal H}}
\newcommand{\ii}{{\mathbf I}}

\DeclareMathOperator{\logb}{\log_2}

\DeclareMathOperator{\prob}{\mbox{Prob }}
\DeclareMathOperator{\proj}{\mbox{Proj}}
\DeclareMathOperator{\tr}{\mbox{Tr}}

\newcommand{\context}{C}
\newcommand{\shentr}[1]{S\left({#1}\right)}

\newcommand{\dasei}[2]{\underline{\delta_{#1}\left(#2 \right)}}
\newcommand{\daseo}[2]{\overline{\delta_{#1}\left(#2\right)}}
\newcommand{\pcl}[1]{\lnot {#1}}
\newcommand{\xcl}[1]{\overline{{#1}\vphantom{{#1}^\ast}}}

\title{Quantifying daseinisation using Shannon entropy}

\author{Rom\`an Zapatrin}

\begin{document}

\maketitle
\begin{abstract}
Topos formalism for quantum mechanics is interpreted in a broader, information retrieval, perspective. Contexts, its basic components, are treated as sources of information. Their interplay, called daseinisation, defined in purely logical terms, is reformulated in terms of two relations: exclusion and preclusion of queries. Then, broadening these options, daseinisation becomes a characteristic of proximity of contexts; to quantify it numerically, Shannon entropy is used. 
\end{abstract}

\section{Introduction}

Topos theory has been suggested by D\"oring and Isham \cite{andreas1} as an enriched mathematical structure to incorporate physical theories, first of all, quantum mechanics (the reader is referred to \cite{floribrief} for a brief outline and to \cite{florifull} for a detailed one). The basic notion of the D\"oring-Isham approach is that of context. Context in its initial formulation is a classical snapshot of a possible variety of the outcomes of an experiment. I suggest to view it in a more general perspective: each context is a source of information, making measurement can be viewed as information retrieval, or pattern recognition. This approach is inspired by Melucci metaphor \cite{Melucci2010}, treating information retrieval process as a physical measurement. Within the suggested approach the difference between physical experiment and information retrieval smears away. Furthermore, topos formalism can be treated as a universal operationalistic phenomenological theory in a broader context than just physics. 

The next essential step towards a complete picture was to introduce a kind of interplay between non-comparable contexts, called {\em daseinisation}. In traditional topos formalism, daseinisation is formulated in logical terms: ``IF something true THEN something true'' (for the details see section \ref{sdasein}). The idea of this essay is quantifying daseinisation numerically using Shannon entropy. In its initial form, daseinisation is too rough to decsribe the interplay, for instance, in standard quantum mechanics, a small unitary perturbation may lead to a drastical jump in discrete daseinisation, while using Shannon proximity, this transition remains smooth.

\section{Contexts}\label{scontexts}

In this essay, I dwell on a `lightweight' (although including standard quantum mechanics) version of topos formalism, when contexts are associated with random variables with a finite number of values. Step by step we going to shift from purely physical to a more general operationalistic, phenomenological viewpoint. In particular, when defining the ingredients of a theory, sooner or later certain notions will remain undefined, treated as basic ones. I suggest to leave the notion of context not to be defined (yet, at least) formally. Instead, we focus on
\begin{itemize}
\item What contexts can be represented with, what can serve as proxies of contexts
\item What an abstract researcher can get out of a context
\end{itemize}

The following scenario is meant: a researcher, call him an {\em agent}, makes a query within a given context $\context$ and with certainty receives one (and only one for each query) of the alternative answers. A context can thus be treated as a generalization of measurement apparatus, or, more formally, as an exhaustive collection of alternatives. Examples of contexts representatives are:
\begin{itemize}
\item A device with pointer. Distinct positions of the pointer correspond to the alternatives.
\item A commutative algebra with a partition of its unit element by idempotents. 
\item A configuration space of a classical system is {\em not} a context. Any its partition, including trivial ones, is a context. The elements of the partition are alternatives.
\item A Hilbert state space $\hh$ of a quantum system is {\em not} a context. A partition of the unit operator in $\hh$ by projectors is a context.
\end{itemize}
Next, let us see what are contexts for. From a general perspective, what is received by an agent making a query? The general answer is {\em information}. I deliberately put apart anything related to the meaning of this information, the only thing needed in this formalism is just a possible amount of information. Now we are a position to give the first numerical characteristics of a context. This is the maximal amount of information that can be retrieved during an individual query. To evaluate it, we use Shannon entropy
\[
\shentr{\context}
=
\logb \# \context
\]
where $\# \context$ is the number of alternatives within the context $\context$. For a device with a pointer $\# \context$ is the number of distinct positions of the pointer. 

Two measurement apparata belong to the same context if they are equivalent: when we measure something, we gain information, and if using apparatus $A$ or $B$ can be treated as the same source of information, they belong to the same context. It can happen that context $B$ is a coarse-graining of context $A$. That means, that if we use $A$, we can always know with certainty, what answers could give us queries within $B$, that is $B$ does not add any information with respect to the context $A$. 

When dealing with finite quantum systems with the state space $\hh$ being Hilbert space of finite dimension, queries are associated with projectors in $\hh$. A context in this case is a decomposition of the identity operator~$\ii$
\[
\ii
=
\sum_J P_J
\;,
\mbox{ with }
P_J\,P_K=0 \quad \forall J\neq K
\]
by mutually orthogonal (not necessary one-dimensional) projectors, an example will be provided below. 

For a classical system, a context can be associated with a partition of its configuration space. The configuration space itself is the finest context, all others are coarse-grainings of it. This is not the case for quantum systems, where the finest context does not exist, this is the core statement of Kochen-Specker theorem. 

Now let us move to quantification of contexts. I suggest a straightforward characteristics, namely, the information capacity of a context. Making a query within a context $\context$, the maximal amount of information we can get of it is 
\[
H(\context)
=
\logb \#(\context)
\]
the logarithm (with base 2) of the number of atomic queries in $\context$. However this measure says nothing about the interplay between contexts. To bind contexts, the notion of {\em daseinisation} is introduced.

\section{Daseinisation via exclusion and preclusion}\label{sdasein}

Daseinisation \cite{andreas2} is a way to look on queries from a context $\context_1$ from a perspective of a context $\context_2$. Daseinisation is of purely logical nature, in \cite{andreas2} {\em inner} and {\em outer} daseinisation are introduced. Begin with formal definitions. Let $Q$ be a query from a context $\context_2$. Then its outer daseinisation within the context $\context_1$ is the least query $\daseo{1}{Q}$ such that
\[
\daseo{1}{Q} \mbox{ is true in } \context_1
\Longrightarrow
Q \mbox{ is true in } \context_2
\]
Dually, the inner daseinisation is defined as the greatest query $\dasei{1}{Q}$ such that
\[
Q \mbox{ is true in } \context_2
\Longrightarrow
\dasei{1}{Q} \mbox{ is true in } \context_1
\]
In a degenerate case, when $\context_1$ is a coarse-graining of $\context_2$, any query from $\context_1$ is a query in $\context_2$, therefore 
\[
\daseo{1}{Q}=\dasei{1}{Q}=Q
\]
We are going to reformulate these notions in a more operationalistic terms. For that, we consider two operations on queries. The first is the {\em exclusion}, it acts queries within a given context $\context$ 
\[
Q\mapsto \xcl{Q}
\]
so that $\xcl{Q}$ is the greatest query within $\context$, which is excluded by $Q$, intuitively this is NOT $Q$ within $\context$. The next notion is {\em preclusion}, which brings us from a context $\context_1$ to another context $\context_2$
\[
Q \mapsto \pcl{Q}
\]
so that $\pcl{Q}$ is the greatest query within $\context_2$, which is excluded by $Q$, intuitively this is NOT $Q$ within $\context_2$. Now the outer daseinisations is the preclusion of the exclusion of $Q$
\[
\daseo{1}{Q}
=
\pcl{\xcl{Q}}
\]
while theinner daseinisation is the exclusion of the preclusion of $Q$.
\[
\dasei{1}{Q}
=
\xcl{\pcl{Q}}
\]

\paragraph{How it looks in quantum mechanics.} Within quantum mechanical realm, an element $Q$ of a context is associated with a projector $Q$ in Hilbert state space $\hh$, then the exclusion is the complement of $Q$, that is
\[
\xcl{Q}
=
\ii - Q
\]
where $\ii$ is the unit operator in $\hh$. Preclusion looks slightly more complicated. A context $\context$ is a partition of the unit operator in $\hh$ by projectors
\[
\context
\mapsto
\ii
=
\sum_{P_j\vert P_j P_k = 0} P_j 
\]
then the preclusion of a projector $Q$ within the context $\context$ is
\[
\pcl{Q}
=
\sum_{P_j\vert QP_j=0} P_j
\]
Obviously, if the query $Q$ belongs to context $\context$, its preclusion within $\context$ coincides with its exclusion:
\[
Q \in \context
\;\Rightarrow\;
\pcl{Q}
=
\xcl{Q}
\]

\bigskip

Such definition looks too rough. Let $\hh=\cfield^2$, consider a basis $\{\ee_1,\ee_2\}$ and the context $\context$ associated with appropriate projectors. Consider a family $\context_\alpha$ parametrized by an angular parameter $\alpha$, each associated with a rotation of the initial basis by the angle $\alpha$. Let $Q=\proj(\ee_1)$. Then its daseinisation is trivial, because the preclusion $\pcl{Q}$ is
\[
\pcl{Q}
=
\left\lbrace
\begin{array}{l}
Q\mbox{, if $\alpha=0$}
\\
0\mbox{ otherwise}
\end{array}
\right.
\]
so, a more subtle measure of proximity is needed.

\section{Measuring proximity by conditional entropy}\label{sproxymity}

Given two contexts $\context_1$ and $\context_2$, each of them can be treated as a source of random variables. Suppose a measurement is performed within the context $\context_2$, then the the next one is carried out within the context $\context_1$, and the agent is aware of the previous result\footnote{Note that I am {\em not} speaking about different contexts related to the same physical object. Furthermore, within the suggested approach the very notion of a physical object becomes secondary. We just have observers-agents, who exchange information and their queries.}. Above the exclusion and preclusion were introduced as logical operations on queries. However, each such operation can be treated as gaining information: to exclude something is to tell something. Let us make the next step and see the things broader, instead of preclusion we just speak about gaining information. If $\pcl{Q}=\context_2$, that means that nothing can be said within context $\context_2$ even if a query $Q$ was made within the context $\context_1$.

To evaluate the added amount of information, the conditional entropy is used. Recall that a pair of random variables $A,B$ can be characterized by conditional entropy
\[
\shentr{A\vert B}
=
\sum_J
B_J
\shentr{A\vert B_J}
=
\sum_J
B_J
\sum_K
A_{(K\vert J)}
\logb A_{(K\vert J)}
\]

\paragraph*{Example.} In classical case, two different contexts are two different partitions of the configuration space. 

\paragraph*{Example.} In quantum realm, queries are associated with projectors and contexts are complete sets of mutually orthogonal projectors. Given a projector $A$ from $\context_1$ and a projector $B$ from $\context_2$, the conditional probability $\prob(A\vert B)$ is
\begin{equation}\label{econdentq}
\prob(A\vert B)
=
\frac{\tr (AB)}{\tr B}
\end{equation}

Now, when all the definitions are set, we pass to the main point. Suppose we have two contexts $\context_1$, $\context_2$. Once we fix a query $B$ from $\context_2$, it gives rise to a probability distribution $P_B(A)$ on the set of queries of $\context_1$, namely, the conditional probabilities 
\[
P_B (A)
=
\prob(A\vert B)
\]
For this distribution we can calculate its entropy, which is just a function $S(A)$ taking numerical values on the elements (=queries) of $\context_1$:
\[
S(A)
=
-\sum_K
A_{(K\vert J)}
\logb A_{(K\vert J)}
\]

\medskip

The first consequence is that if $\context_1$ is a coarse-graining of the context $\context_2$, then no measurement within $\context_1$ brings us any new information if we know the result of $\context_2$, that is $S(A)=0$ for all $A$. This fact has a fundamental meaning: mutual information can be used as a definition of coarse-graining!

\paragraph*{An example.} Continue the example of a quantum system with two degrees of freedom and two contexts associated with the bases in $\hh$:
\[
\context_1\mapsto\{\ee_1,\ee_2\} 
\quad ; \quad
\context_2\mapsto\{\ff_1,\ff_2\} 
\]
Denote $\alpha$ as
\[
\cos \alpha
=
\left\lvert\left(
\ee_1,\ff_1
\right)\right\rvert^2
\]
Then the usual daseinisation of any query becomes trivial as $\alpha\neq 0, \pi/2$, while Shannon proximity is more flexible, namely
\[
\shentr{\context_1 \vert \context_2}
=
-\cos^2\alpha \logb\cos^@ \alpha
-
\sin^2\alpha \logb\sin^2\alpha
\]
Now, let $\alpha\simeq 0$ is small, then given a query related to he projector $\proj_{\ff_1}$ within $\context_2$ we can confer its daseinisation in $\context_1$ which is anyway
\[
\daseo{1}{\proj_{\ff_1}}
=
\left\lbrace
\begin{array}{l}
\proj_{\ff_1}\mbox{, if $\alpha=0$}
\\
\ii\mbox{ otherwise}
\end{array}
\right.
\]
where $\ii$ stands for the unit operator in $\hh$, while the proximity of contexts reads
\[
\shentr{\context_1 \vert \context_2}
\simeq
\alpha^2
\]
thus capturing the proximity of the contexts.

\section*{Concluding remarks}

Starting from the idea that physical measurement can be viewed as information retrieval, we consider contexts as sources of information. Next, we study the interplay between contexts in purely information terms. In particular, instead of defining ab initio the context category, we bring it on an operationalistic basis, namely, if we perform something within context $\context_2$ and then we can with certainty predict the result within the context $\context_1$, that is, $\context_1$ provides us with no additional information with respect to context $\context_2$, we say, that there is an arrow $\context_2 \to \context_1$, so the set of available contexts is given a structure of category. 

Another consequence is that within this information-theoretical approach the notion of a physical system smears. We are just speaking about observers who perform queries and exchange information. In this terms any collection of contexts may be said to be a system, and the only meaningful thing remains to speak about inclusion of systems, that is, about subsystems.

\end{document}